\documentstyle[11pt,mrs2003,epsfig]{article}
\begin{document} 
\title{PROJECT OF A SUPERFLUID $^3$HE DETECTOR FOR DIRECT DETECTION OF NON-BARYONIC DARK MATTER~:~MACHE3}

\author{E. MOULIN, F. NARAGHI, D. SANTOS}
\affil{Laboratoire de Physique Subatomique et de Cosmologie, 
 CNRS/IN2P3 and Universit\'e Joseph Fourier, 
53, avenue des Martyrs, 38026 Grenoble cedex, France} 
\author{E. COLLIN, C. WINKELMANN, Y. BUNKOV, H. GODFRIN}
\affil{Centre de Recherches sur les Tr\`es Basses Temp\'eratures,
 CNRS and Universit\'e Joseph Fourier, BP166, 38042 Grenoble cedex 9, France} 
 
\begin{abstract} 
MACHe3 (MAtrix of Cells of superfluid Helium 3) is a project of non-baryonic Dark Matter search using superfluid 
{\mbox{$ ^{3}{\mathrm{He}}$}} as sensitive medium. Simulations on a high granularity matrix show very good rejection 
against background events.
First results on a prototype cell are very encouraging.
Neutron detection has been 
highlighted as well as cosmic muon detection. A phenomenological study has been done with the DarkSUSY code to investigate 
complementarity of MACHe3 with existing Dark Matter detectors.
\end{abstract}  
\section{Introduction} 
The estimation of cosmological parameters 
has reached an unprecedented level of accuracy since the new data on the cosmic microwave background \cite{wmap},\cite{archeops} 
used in combination with other cosmological observations, such as the SNIa and galaxy cluster surveys.  
Recent results show that matter contribution to the total density of the Universe is mainly composed of non-baryonic dark matter.
Supersymmetric
theories generally predict this yet still not observed and mostly non-interacting matter to be composed of WIMPs, which best motivated
candidate is the neutralino \cite{jungman}. 

Many collaborations have developped promising detectors to hunt actively non-baryonic dark matter
candidates. These new detectors have reached significant sensitivity to begin to test regions of the SUSY parameter
space compatible with accelerator and cosmological constraints   
and improved the upper limits on scalar and 
axial interaction exclusion plots \cite{edelweiss}. 
Direct detection experiments present common problems such as neutron interaction and radioactive contamination of both sensitive medium
and surrounding materials.
Since early experimental works \cite{lanc},\cite{nature95},\cite{prb98}, a superfluid {\mbox{$ ^{3}{\mathrm{He}}$}} detector has been 
proposed \cite{nima} for direct detection of non-baryonic dark matter search.
Monte Carlo simulations have 
shown that a high granularity detector, a matrix of superfluid {\mbox{$ ^{3}{\mathrm{He}}$}} cells, would allow to reach a high rejection factor
against background events, leading to a low false neutralino event rate. 
\section{MACHe3 project} 
\subsection{Superfluid {\mbox{$ ^{3}{\mathrm{He}}$}}-B as a privileged sensitive medium }
At ultra-low temperature ($\simeq$ 100 $\mathrm{\mu}$K), {\mbox{$ ^{3}{\mathrm{He}}$}} is a very interesting medium since it 
could be used as a highly sensitive bolometer.
As a matter of fact, around 100 $\mathrm{\mu}$K and at low pressure, {\mbox{$ ^{3}{\mathrm{He}}$}} is in its superfluid B phase 
which has an extremely small quasiparticle gap.
The use of superfluid {\mbox{$ ^{3}{\mathrm{He}}$}}-B 
is motivated by its very appealing features compared to other materials~:\\
\begin{itemize}
\item {\mbox{$ ^{3}{\mathrm{He}}$}~} being a 1/2 spin nucleus, 
will be mainly sensitive  
to axial interaction, making this detector complementary to existing ones, mainly sensitive to the scalar interaction. The axial
interaction is largely dominant in all the SUSY region associated 
with a substantial elastic cross-section \cite{plb}.
\item  An ultra-high purity due to its superfluid phase, only {\mbox{$ ^{4}{\mathrm{He}}$}~} can be dissolved in superfluid {\mbox{$ ^{3}{\mathrm{He}}$}~} but 
in an extremely low fraction \cite{prb98}.
\item  A high neutron capture cross-section ($\sim$10 barns for 10 keV neutrons), implying a clear signature well discriminated from a WIMP signal.
\item  Low Compton cross-section. No intrinsic X-rays.
\item  A low detection threshold, less than 1 keV.
\item  A high signal to noise ratio, due to the small energy range expected for a WIMP signal.
The maximum recoil energy does only slightly depend on the
WIMP mass, given that the target nucleus ($\rm m_{\,^3He}\,=\,2.81\,GeV\,/c^2$) is much
lighter than the incoming $\tilde{\chi}$ (${\rm M_\chi \geq 32 \,GeV}\,/c^2$). As a matter of fact, the recoil
energy range needs to be studied only below ${\rm 6 \,keV}$, see \cite{nima}. 
\end{itemize}
\subsection{A high granularity matrix}
\begin{figure}  
\vspace*{1.25cm}  
\begin{center}
\epsfig{figure=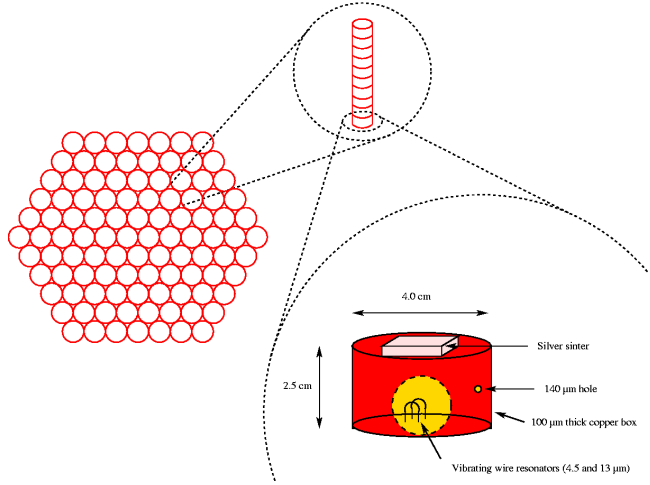,width=8.5cm}  
\end{center}
\vspace*{0.25cm}  
\caption{A top view of the hexagonal matrix. Each cylinder is divided in 10 Lancaster type bolometer {\mbox{$ ^{3}{\mathrm{He}}$}~}-cells.
Each cell is made of copper and filled with superfluid {\mbox{$ ^{3}{\mathrm{He}}$}~}. A silver sinter piece is on the top for cooling purpose. Two NbTi vibrating wire resonators 
(VWR) forming semi-loop on legs are inside, of 4.5 and 13 $\mu$m diameter respectively.} 
\label{fig:scheme}
\end{figure}

\begin{figure}[!h]  
\vspace*{1.25cm}  
\begin{center}
\epsfig{figure=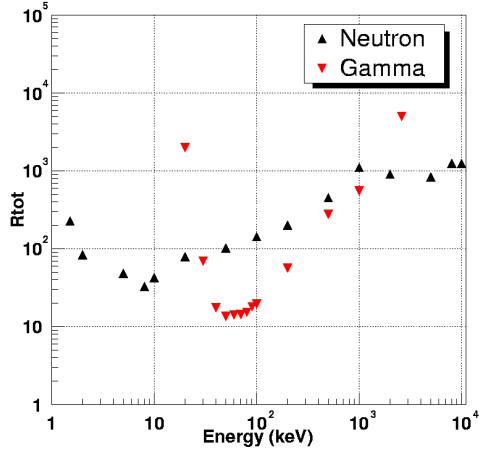,width=6.5cm}  
\end{center}
\vspace*{0.25cm}  
\caption{Total rejection versus incoming kinetic energy particle, for $\gamma$-rays and neutrons respectively.
The total rejection is defined as the number of incident particles over the number of false $\tilde{\chi}$~ event.
99.98\% of 2.6 $\gamma$-rays are rejected. For 1 MeV neutrons, the total rejection is 99,90\%.} 
\label{fig:rejection}
\end{figure}

MACHe3 project will consist of a 10 kg matrix composed 
of 1000 cells of 125 cm$^3$ filled with superfluid {\mbox{$ ^{3}{\mathrm{He}}$}} at 100 $\mu$K. A privileged design is shown on figure~\ref{fig:scheme}. It is based on
a hexagonal matrix with a close to a bee nest configuration. Each cylinder is divided in 10 Lancaster type \cite{lanc} cylindric bolometer cells.     
Rejection principle relies on the energy release measurement and the correlation
among the cells to discriminate
background events (neutrons, $\gamma$-rays and muons) from neutralino signal.
It allows to 
reject background events given the fact that a WIMP fires only one cell with an energy release below 6 keV. 
The high granularity combined with energy deposit allows to obtain relatively good rejection as seen
on figure~\ref{fig:rejection}.
For 2.6 MeV gammas, only one over 5000 mimicks a WIMP event, meaning that 99.98\% are rejected. 
Concerning neutrons, we benefit from the high neutron capture cross-section.
The capture process, $\mathrm{n\,+\,{\mbox{$ ^{3}{\mathrm{He}}$}~}\,\rightarrow\,p\,+\,^3H\,+764\,KeV}$, releases 764 keV in one cell.
It allows to reach a very good rejection, 99.90\% of 1 MeV neutrons are rejected \cite{moulin}.
In underground environnement, the false event rate is estimated to be less than $\sim$10$^{-1}$ day$^{-1}$
for neutrons and $\sim$10$^{-2}$ day$^{-1}$ for muons \cite{nima}.  
\subsection{Experimental results on a prototype cell}  
The elementary component of MACHe3 is the superfluid {\mbox{$ ^{3}{\mathrm{He}}$}} cell \cite{prb98}.
An experimental test of a small prototype cell has been done with a lead shielding 
around the cryostat to estimate the contribution of the external
natural radioactivity. 
The cell was a copper cylindric box 
($\mathrm{V} \simeq 75 \,{\rm mm}^3$) filled with superfluid {\mbox{$ ^{3}{\mathrm{He}}$}} around 100 $\mathrm{\mu}K$. The detection principle relies on the damping 
on the vibrating wire resonance. More precisely, the vibration amplitude decreases when a particle interacts inside the cell. 
The thermal diffusion is done through the hole. See \cite{nature95} for more details.
\begin{figure}[!hb]  
\vspace*{0.0cm}  
\begin{center}
\epsfig{figure=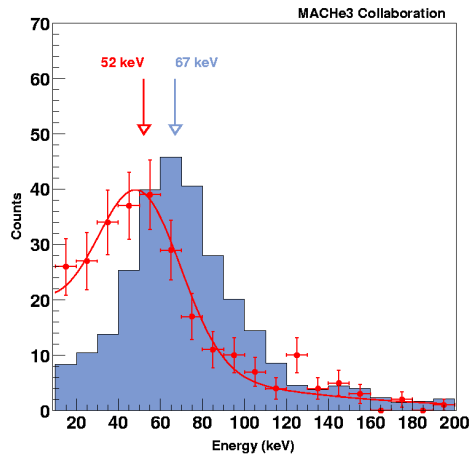,width=6.5cm}  
\end{center}
\vspace*{0.25cm}  
\caption{Muon detection with the prototype cell. Experimental data (points) exhibit a peak at 52 keV.
The result of the Geant4 simulation (histogram) which describes the main features of the set-up, shows a peak at 67 keV.} 
\label{fig:detection}
\end{figure} 

Stability in temperature for one week per cycle at T $\mathrm{\simeq}\,100\,\mathrm{\mu}K$
and detection threshold around 1 keV have been obtained.
An experiment with an Am-Be neutron source has been performed to validate the response of the
cell to neutron irradiation. The capture has been observed \cite{collin}. A 14\% of the total energy released by the capture is lost in scintillation and 
vortex production. To disentangle these effects, development of a new prototype cell 
with an additional signal from the temperature change of the cell walls is in progress. 
The detection of cosmic muons during 12 hours has been highlighted as displayed on figure~\ref{fig:detection}. 
Experimental data exhibits a peak at 52 keV.
A Geant4 \cite{g4} simulation reproducing the experimental set-up has been developped
to estimate the released energy by 2 GeV muons crossing through the cell. 
Simulation presents a clear peak at 67 keV. A shift compared with expectation value is
observed. The energy loss may be related to UV photon emission and vortex formation \cite{collin}.
\section{Phenomenological study}
A phenomenological study has been done with the DarkSUSY code\footnote{The version used is 3.14.01, 
with correction of some minor bugs.} \cite{ds}, to test performance of MACHe3 and its complementarity with existing detectors within
the framework of a phenomenological supersymmetric model. A scan on a large range of all free parameters has been done, 
corresponding to a total number of $\sim 2 \times 10^6$ models. 
Accelerator constraints as well as cosmological ones have been considered \cite{plb}.
\begin{figure}[!hb]  
\vspace*{1.25cm}  
\begin{center}
\epsfig{figure=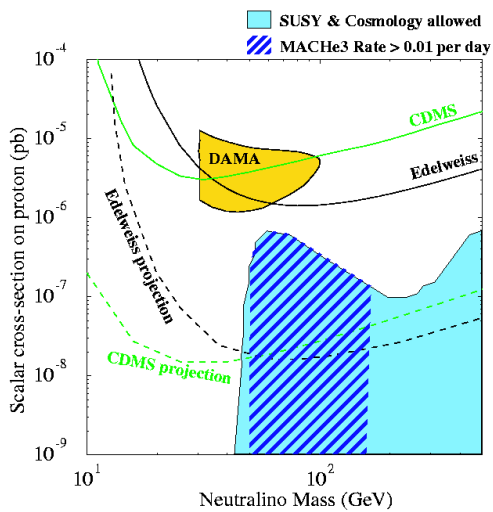,width=6.5cm}  
\end{center}
\vspace*{0.25cm}  
\caption{Scalar cross-section (on proton) as a function of the neutralino 
mass. Exclusion limits from the Edelweiss \cite{edelweiss} and CDMS \cite{cdms} experiments are 
shown, as well as the $3\,\sigma$ DAMA region \cite{dama}.  
Dashed lines indicate projected limits from Edelweiss and CDMS. 
Light blue points indicate models allowed by accelerator constraints and cosmology.
The blue-hased region corresponds to models giving a 
neutralino rate in MACHe3 higher  
than the estimated background level of $10^{-2}\,{\rm day}^{-1}$.} 
\label{fig:susy}
\end{figure} 
Using the DarkSUSY code, the $\tilde{\chi}$~-{\mbox{$ ^{3}{\mathrm{He}}$}~} axial
cross-section has been evaluated. 
The $\tilde{\chi}$~ event rate has then been evaluated for a $10\,{\rm kg}$ {\mbox{$ ^{3}{\mathrm{He}}$}} matrix, 
to be compared with the background rate \cite{nima}.
A large number of models are giving a rate higher than the estimated false event rate induced by neutrons 
($\sim 10^{-1}\;\mathrm{day}^{-1}$), or above the estimated muon background ($\sim 10^{-2}\;\mathrm{day}^{-1}$) \cite{plb}.
The $\mu$ background level ($10^{-2}\;\mathrm{day}^{-1}$) is chosen as the lowest reachable limit
for MACHe3. Models giving a rate higher than the $\mu$ background are selected. Figure~\ref{fig:susy}
shows how they are distributed in the scalar cross section on proton neutralino mass diagram. DAMA region is displayed as well as 
exclusion curves from Edelweiss and CDMS and their projections.
It can be seen that many SUSY models lie below the projected limits of future scalar detectors 
(Edelweiss \cite{edelweiss}, CDMS \cite{cdms}), while giving an event rate above 
the value taken as the lowest reachable limit for MACHe3.
\begin{figure}[!h]  
\vspace*{1.25cm}  
\begin{center}
\epsfig{figure=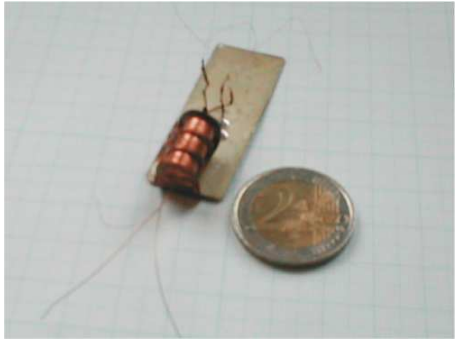,width=6.0cm}  
\end{center}
\vspace*{0.25cm}  
\caption{A picture of the multicell prototype. It is composed of 3 cylinder cells. Vibrating wires of 4.5 $\mu$m
diameter are visible. Middle cell walls are common with the two other neighbouring cells.} 
\label{fig:prototype}
\end{figure} 
\section{Conclusions}
Detection principle and cosmic muon detection in the elementary component of our matrix has been shown.
Background rejection has been presented.
It has been highlighted that a ${\rm 10\,kg}$ high granularity {\mbox{$ ^{3}{\mathrm{He}}$}} detector (MACHe3) 
would allow to obtain, in many SUSY models, 
a $\tilde{\chi}$ event rate higher than the estimated background. 
MACHe3 would thus 
potentially allow to reach a large part of the SUSY region, not excluded by current collider limits and 
projection limit of runing detectors.
An experiment on a multicell prototype as seen on figure~\ref{fig:prototype}
devoted to validate correlation among the cells 
for background rejection is in progress.
%
%
 
\vfill 
\end{document}